# Modeling of Water Evaporation in Hydrogels from Aspect of Mechanical Analytics

Zehua Yu[1], Yongshun Ren[1], Kang Liu[1, †]

[1] MOE Key Laboratory of Hydraulic Machinery Transients, School of Power and Mechanical Engineering, Wuhan University, Wuhan 430072, China

[†] kang.liu@whu.edu.cn

**Abstract**

Water evaporation is critically important for hydrogels in open-air applications, but theoretically modeling is difficult due to the complicated intermolecular interactions and sustained deformation. In this work, we construct a simplified model to describe the state of water inside the hydrogel by only considering mechanical stretching. We employ "negative pressure" to bridge the stretching force in water and elastic force generated by the polymer network. Combined with a constitutive equation of elasticity for hydrogels and classic diffusion equation, this model gives a universal solution to calculate the saturated vapor pressure, dynamic evaporation rates and real-time deformation of different hydrogels. The calculated results agree well with experiments results both in steady state and dynamic process for commonly used poly 2-hydroxyethyl methacrylate and polyacrylamide hydrogels with diverse components. In addition, the model predicts that, hydrogels with high modulus shows stronger ability to retain water in open environment.

## 1 Introduction

Hydrogels are three-dimensional cross-linked networks of hydrophilic polymer chains with water filling the interstitial spaces.[1] The high water content, flexibility and compatibility with human tissues make hydrogels widely utilized in wound dressings,[2,3] implantable devices,[4-6] and related bioengineering fields.[7,8] Additionally, with physical properties similar to solids and plentiful water inside, hydrogels also attract increasing attention in applications of energy conversion and thermal management.[9-13] Among the expanding applications, hydrogels are frequently utilized in open-air environments rather than in aqueous liquids. In these scenarios, water inside hydrogels is prone to evaporate, leading to dehydration of hydrogels,[14] which underscores the importance of understanding the evaporation of water in hydrogels. During the water evaporation, polymer network of the hydrogel undergoes sustained shrinkage. Interaction between polymer chains and water molecules and interaction among water molecules varies concurrently.[15,16] Difficulties in quantitative description of these microscopic interactions lead to challenges in constructing a theoretical model to characterize the evaporation behavior of water in hydrogels.[17]

Flory and Huggins developed a basic theoretical framework to analyze the thermodynamics in polymer solutions,[18-20] which has also been employed to investigate hydrogels after integrating an elastic term (Flory-Rehner theory),[21] especially in describing the swelling of hydrogels in aqueous solutions.[22,23] However, Flory-based theory is initially constructed based on a basic assumption of randomly dispersed polymer solution,[18] and describe the interaction between polymer chains and solvent molecules with mixing entropy and mixing enthalpy. In hydrogels, the polymer chains are polymerized as a network with pores of several nanometers in diameter.[24,25] The network deforms but can hardly dislocate. The pores separate the water into water pockets.[26-28] During dehydration of hydrogels in air, water becomes to be a small proportion inside the hydrogel. Rigidity of the polymer network starts to generate a force to stop the volume shrinkage due to evaporation and induce mechanical stretching in water. Chemical potential of water dramatically decreases, the same mechanism as water inside capillary pores.[29,30] Structurally, hydrogels are more like deformable pores with water inside than solution.

Based on the analysis above, we construct a model that consider the hydrogel as a deformable matrix with nano-sized pores. Mixing entropy and mixing enthalpy from polymer-water interaction are neglected. Deformation of the hydrogel is considered to be caused by cohesive force induced mechanical stretching in water, and degree of the deformation is determined by the balance between mechanical stretching in water and rigidity of the polymer network. We employed the concept of negative pressure to describe the stretching in water and also as a bridge to establish mechanical equilibrium with the stress of the hydrogel. Combined with a constitutive equation of elasticity for hydrogels and classic diffusion equation, we build a universal model to calculate the saturated vapor pressure, dynamic evaporation rates and deformation of different hydrogels. We compared the theoretical results with experiment results. The calculated results agree well with experiments results both in steady state and dynamic process for commonly used hydrogels, showing applicability of the model in some engineering applications.

## 2 Results and discussion

### 2.1 Mathematic model of water evaporation in hydrogels

To show the evaporation behavior of water in hydrogels, we monitored the mass variation of two common hydrogel materials, poly(2-hydroxyethyl methacrylate) (PHEMA) and polyacrylamide (PAAM), exposed in environment (fabrication can be found in supplementary information note S1). In the experiment, hydrogels were cut into thin slices with diameters of approximately 5 cm. The thickness was controlled to be 50-100 μm after swelling. This ensured that the internal mass transfer resistance was negligible, and the dehydration of hydrogels could be considered as one-dimensional. As a comparison, the evaporation of pure water was observed by pouring water into a Petri dish with a diameter of 34.4 mm. A balance was employed to record the mass variation of the hydrogels and pure water. The ambient temperature is 25 °C, and humidity is 80% RH. As depicted in Fig. 1a, the mass of pure water shows a linear decrease over time until the water is completely evaporated. While the relative mass ($m_{wc} / m_{wo}$) of water in hydrogels shows a nonlinear variation over time. The slope of the curve represents the mass change rate of water in hydrogels due to evaporation, and the rate gradually decreases and approaches zero in equilibrium. In steady-state, both the PHEMA and PAAM hydrogels retained a certain amount of water. From the mass flux of water vapor ($J_v$), we can calculate the vapor pressure ($p_v$) according to a simple diffusion equation

$$J_v = k_1(p_{v,sat} - p_v) \tag{1}$$

$p_{v,sat}$ is the saturated vapor pressure, $k_1$ is the mass transfer coefficient, which is taken as $1.6 \times 10^{-5}$ g s$^{-1}$ m$^{-2}$ Pa$^{-1}$ according to our experiment of pure water evaporation.

The saturated vapor pressure of pure water remains constant during the experiment. While both hydrogels demonstrate a varying saturated vapor pressure during dehydration (Fig. 1b). For PAAM, the saturated vapor pressure during the initial stage is nearly the same as that of pure water, because of the high volume fraction of water (96 vol%). However, as evaporation progresses, the vapor pressure of PAAM begins to decline. This decline continues until the vapor pressure equals to the ambient vapor pressure, and evaporation stops. PHEMA hydrogel exhibits the ability to regulate vapor pressure from the beginning of evaporation, due to the low volume fraction of water (37 vol%). As evaporation progresses, the vapor pressure declines to the same value as that of PAAM hydrogel. We tried to use Flory-Rehner theory to explain the varying vapor pressure of the two hydrogels.[21] The chemical potential ($\mu_w$) of water inside the hydrogel is calculated as

$$\mu_w(T,p) = \mu_w^o(T,p_e) + RT\left[\ln(1-\varphi_w) + \varphi_w + \chi\varphi_w^2\right] + RTK\left(\varphi_w^{1/3} - \varphi_w/2\right) \quad (2)$$

Where $T$ and $p$ are temperature and pressure, respectively. $R$ is gas constant. $p_e$ is ambient pressure. $\varphi_w$ is the water volume fraction, $\chi$ is the polymer-solvent interaction coefficient. $K$ can be calculated as $K = v_e v_1 / V_o$, $v_e$ is the effective number of crosslinking units, $v_1$ is the volume of a single solvent molecule, and $V_o$ is the initial dry volume of the polymer. The second term on the right side of the equation represents the mixing term (arises from interactions between polymer and water) and the third term denotes the elastic term (originates from the configurational entropy of the polymer under varying swelling conditions). Through a transformer of Kelvin equation, chemical potential can be converted into vapor pressure[31,32]

$$p_v = p_{v,sat} \cdot \exp(\mu_w(T,p)/RT) \quad (3)$$

The vapor pressure predicted by the Flory-Rehner model is lower than the experimental results, especially at low water content (Fig. 1b). This discrepancy indicates that chemical potential of water predicted by the Flory-Rehner model is smaller than the experimental value. In Flory-Rehner model, the chemical potential of water is derived from the mixing term and the elastic term.[21] Compared with mixing term, the elastic term is negligible (Fig. 1c). The mixing term can be further divided into mixing entropy and mixing enthalpy. The signs of mixing entropy and mixing enthalpy are opposite, and the magnitude of the former is much greater than the latter at lower water contents (Fig. 1d). Therefore, it can be inferred that the inaccuracy of the Flory-Rehner model at low water content stems from its overestimation of mixing entropy.

The expression of mixing entropy in Flory-Rehner model is derived based on polymer solutions.[18] However, hydrogels contain numerous cross-linking sites that significantly restrict the mobility of the polymer chains (Fig. 1e-f). The polymer network deforms but can hardly dislocate. The actual mixing entropy in hydrogels is much lower than the value predicted by the polymer solution model. In addition, the hydrogel is more like a solid system than a solution system. The deformation and rigidity of the polymer network would also induce strong interaction between solvent molecules, the same as water inside capillary pores of solid-state porous materials.[29,30] In our previous work, we have demonstrated the dependency of saturated vapor pressure on the mechanical strength of hydrogels.[15] Types of functional groups and density of crosslinking do not significantly affect the saturated pressure, especially for hydrogels with low water content. Hence, the inherent state of water would be more important than polymer-water interaction for the saturated vapor pressure.

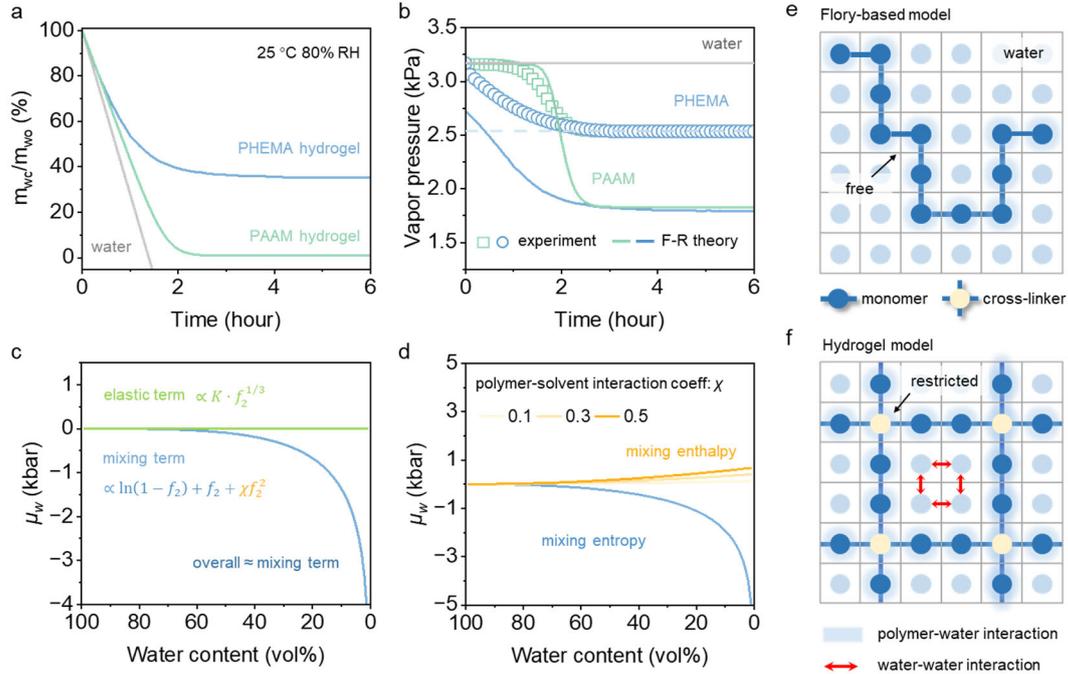

**Figure 1.** Evaporation of water in hydrogels. (a) Relative water mass ($m_{wc}$ / $m_{wo}$) during evaporation for PAAM, PHEMA hydrogels and pure water. $m_{wc}$ represents the current mass of water in hydrogels, $m_{wo}$ indicates the original mass of water in hydrogels (b) Vapor pressure of water in PAAM, PHEMA and pure water from experiments and the predictions from the Flory-Rehner theory (F-R theory). (c) Contributions of the mixing and elastic terms to the chemical potential of water. For both PAAM and PHEMA hydrogels, the polymer-solvent interaction parameter ($\chi$) is set as 0.5. (b) Contributions of the mixing entropy and mixing enthalpy to the chemical potential of water. (e-f) Schematic of the Flory-based model and the hydrogel model in describing entropy term of hydrogels. The light blue shading near the monomers indicates the interaction between the monomers and the solvent molecules.

Based on the analysis above, we establish a model by taking the hydrogels as deformable nanopores. We consider only the mechanical stretching of water and elastic deformation of polymer network for simplify (Fig. 2). The model is one-dimensional considering the bottom surface closely attached on a surface, and the area does no change with time. During evaporation in our model, reduction of water induces volume shrinkage of the hydrogel. At the same time, elastic force of the hydrogel resist shrinks. Balance of the cohesive force in water and elastic force from the polymer network determines the deformation of the hydrogel at equilibrium state. Different from liquid water, the volume shrink ($\Delta l$) is smaller than the reduction of water ($l_w$) due to rigidity of the network. The remaining water within the hydrogel undergoes a stretching state and generate strong cohesive force in water. To establish quantitative connection between the stretching and deformation, we employed "negative pressure", a metastable state in liquid water,[15,16,32] to describe the state of stretching and also calculate the deformation. Since the modulus of the hydrogel is closed related with the water content, mechanic equilibrium on the hydrogel can be derived as:

$$p_{l.gel}(\varepsilon_s) - p_e = -\int_0^{\Delta l} E(\varphi_w) \cdot dl / l_{wet} \qquad (4)$$

$\varepsilon_s$ is the stretching ratio, which can be calculated by: $\varepsilon_s = (1 - \Delta l / l_{w,o}) / \omega$. Here $l_{w,o}$ is calculated as $l_{wet} - l_{dry}$, which is the volume difference between fully swollen hydrogel and fully dried hydrogel. $\omega$ is the relative water content, and can be calculated as $(m_t - m_{dry}) / (m_{wet} - m_{dry})$. $p_{l,gel}(\varepsilon_s)$ is the negative pressure of water under the stretching ratio of $\varepsilon_s$, which is obtained by fitting the results from molecular dynamics simulation (Fig. S1),[33] $\varphi_w$ is the volume fraction of water in the hydrogel, $l_{wet}$ is the volume of fully swollen hydrogel. In equation (4), the elastic force is calculated by integrating the product of the modulus and the strain, because the modulus varies with water contents. Details about the model can be found in Supplementary information note S3.

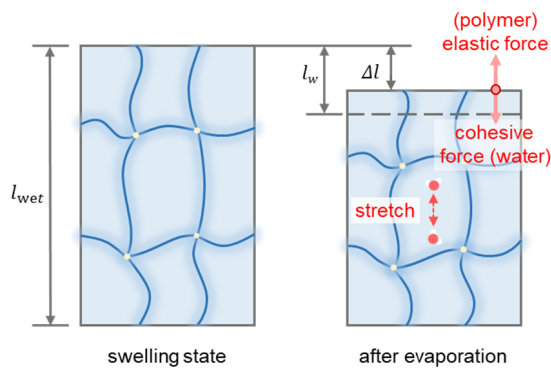

**Figure 2.** Model of water evaporation in hydrogels. Assuming the model is one-dimensional, $l_{wet}$ for the original volume of the hydrogel, $l_w$ for the volume of evaporated water, $\Delta l$ for the actual

volume variation of the hydrogels during evaporation.

The only parameter in equation (4) yet to be determined is the modulus of the hydrogel. We measured the elastic modulus of hydrogels at various water contents (Fig. 3a, S2 and S3, Supplementary information note S2). Interestingly, despite differences in monomer types, concentrations, and crosslinker concentrations between PAAM and PHEMA hydrogels, the elastic modulus at different water contents can be approximately unified onto a single curve. Attempts to fit the modulus data using the classic Flory-Rehner statistical model and simple affine network model (Fig. S4 and S5) showed that these two models perform well only when the water content is higher than 40 vol%.[34,35] They failed to capture the rapid increase of modulus observed at low water contents. The sharp rise in modulus resembles the glass transition behavior in polymers, indicating a great reduction in the degrees of freedom within the polymer network.[36] For hydrogels, we hypothesize that, the sharp rise could be caused by the mass variation of bound/intermediate water (water molecules strongly interacted with polymer via hydrogen bonds) and free water (water molecules can freely move within the hydrogel matrix). The modulus of free water is negligible, while bound/intermediate water form stronger hydrogen bonds with the polymer chains,[37,38] and have high modulus. Free water evaporates before bound/intermediate water. The overall modulus dramatically increases when free water is exhausted.

To demonstrate the hypothesis, Raman spectroscopy was performed to characterize the content of bound/intermediate water and free water during evaporation. The proportion of bound/intermediate water increases sharply from the point of about 30 vol% (Fig. S6).[15,38] This trend aligns with the rapid increase in hydrogel modulus observed in Fig. 3a. These finding suggests a strong correlation between the fraction of bound/intermediate water and the overall modulus of hydrogels. Based on this observation, we developed a model of elastic modulus of the hydrogels. We assumed that the overall modulus of hydrogels ($E_{gel}$) is resulted from a series combination of contributions from the modulus of polymer network ($E_p$), bound/intermediate water ($E_b$) and free water ($E_f$, Fig. 3b). And the constitutive equation for the elastic modulus of hydrogels is derived as (Supplementary information note S4),

$$E_{gel} = \cfrac{1}{\cfrac{\varphi_p}{E_p} + \cfrac{\varphi_b}{E_b} + \cfrac{\varphi_f}{E_f}} \qquad (5)$$

$\varphi$ is the volume fraction. The subscripts *gel*, *p*, *b*, and *f* refer to hydrogel, polymer network, bound/intermediate water, and free water, respectively. $E_p$ and $E_f$ is specified as 4.95 GPa and 16.0 kPa, which are corresponds to the modulus of the completely dry and swollen hydrogels, respectively. In bound/intermediate water, the closer water molecules are to the polymer chains, the more stable their structure is, which produces higher elastic modulus (Fig. 3b). During dehydration, intermediate water evaporates earlier than bound water, leading to changes in both total amount and relative proportion of the two types of water, and consequently alters the $E_b$. Therefore, variation of $E_b$ plays a crucial role in the transition of the hydrogel from a rigid polymer state to a flexible gel state. Accordingly, we define the $E_b$ as a function of $\varphi_b$, and is expressed as $E_b(\varphi_b) = E_{b,o} \cdot \exp(\alpha \cdot \varphi_b)$. $E_{b,o}$ is the modulus of pure bound water, and is determined as 3.2 GPa according to the modulus of ice due to the similar hydrogen bond strength.[39] $\alpha$ indicates the increasing rates of modulus with water contents, which is determined through fitting the results of the experiments (determined as 18.6 for the normalized curve in Fig. 3a). Notably, $E_b$ comes to be a fixed value after the appearance of free water. The constitutive equation fits well with the experimental results (Fig. 3a and S7a).

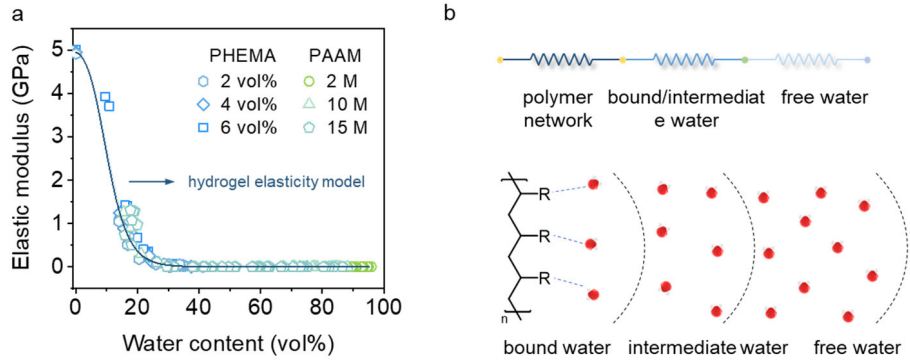

**Figure 3.** Hydrogel elasticity model. (a) Elastic modulus of hydrogels under different water contents. (b) Schematic of the hydrogel elasticity model.

## 2.2 Evaporation behavior of water in hydrogels predicted by the model

With equation (5), we can solve the equation (4) to determine the negative pressure in water and deformation of hydrogels during evaporation (Fig. S8). Combined with equation (1) and (3),[31] we can further derive the steady-state saturated vapor pressure and real-time evaporation rate. As shown in Fig. 4a, variation of vapor pressure with water content depicts a normalized trend for different hydrogels. This trend mirrors the variation of elastic modulus to water content in hydrogels, with a sharp transition at the water content range of 20-40 vol%. Our model effectively captures this trend and accurately predicts the corresponding vapor pressures (Fig. 4a and S7b). Moreover, our model can also show dynamic process of evaporation for hydrogels. The theoretical results agree well with experiment results as shown in Fig. 4b. For PHEMA hydrogel, the evaporation rate in the initial state is a little faster than experiment results, because the model neglect the mass transfer resistance of water in PHEMA. The equilibrium water content is also slightly higher, which is consistent with the discrepancies observed in the vapor pressure prediction.

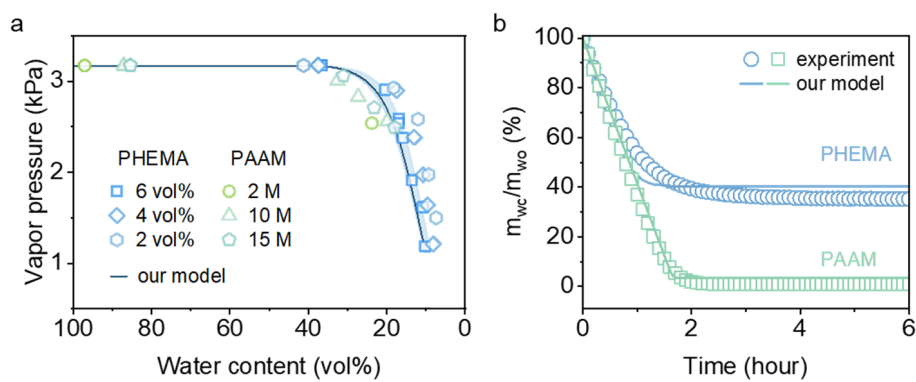

**Figure 4.** Evaporation behavior of water predicted by the model. (a) Vapor pressure of hydrogels under different water contents. (b) Dynamic evaporation process for PHEMA and PAAM hydrogels.

PHEMA and PAAM hydrogels exhibit normalized vapor pressure with water content, because of the normalized changes of modulus with water content. It means, PHEMA and PAAM hydrogel have the same modulus at certain water content, without effect of functional groups and crosslinks. Skeleton of PHEMA and PAAM hydrogels are both constructed by linear carbon-carbon single bonds. But apparently, not all hydrogels show the same trend as PHEMA and PAAM hydrogels, because $E_p$ of hydrogels is different in Equation (5). According to Li et al.,[35] conventionally synthesized PAAM and PHEMA hydrogels can be categorized as elastic hydrogels. Enhancing the structure by introducing a double network (viscoelastic hydrogels) or reinforcing the hydrogel framework to form high-order structure hydrogels significantly increase the elastic modulus. By fitting the modulus of these hydrogels using our elastic constitutive equation (Fig. 5a) and integrating it into the evaporation model, we can predict the dynamic evaporation process of these hydrogels (Fig. 5b). The results reveal that as the modulus increases, the water retention capacity improves. Enhancing the elastic modulus of hydrogels through material design could be a promising strategy for improving water retention under ambient conditions for applications in outdoor conditions.

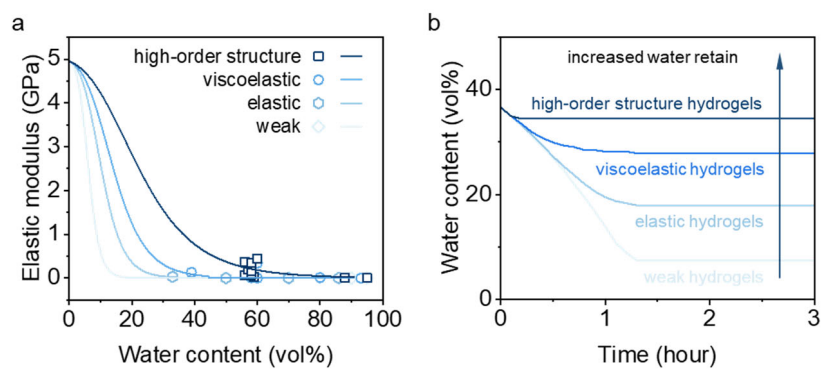

**Figure 5.** Evaporation behavior of different types of hydrogels predicted by our model. (a) Modulus of different type hydrogels. (b) Dehydration of the hydrogels at 25 °C and 80% RH.

## 3  Conclusions

In summary, we have developed a simplified model based on mechanical stretching of water to describe water evaporation in various hydrogels. With the concept of "negative pressure", the model bridges the water tension with the elastic force of the polymer network. Integrating with elastic constitutive equation and the diffusion equation, the model achieves precise description of the vapor pressure and the dynamic evaporation of water in hydrogels. In addition, the model predicts that, hydrogels with high modulus shows stronger ability to retain water in open environment. These results provide insights into the microscopic mechanism of water evaporation in hydrogels and offers guidance for materials design in different applications.

**Supporting Information**

Additional experimental details, results, characterizations and derivation of the models

**Author Information**

Corresponding Author

Kang Liu – *MOE Key Laboratory of Hydraulic Machinery Transients, School of power and Mechanical Engineering, Wuhan University, Wuhan 430072, China;* orcid.org/0000-0002-2781-2581; Email: kang.liu@whu.edu.cn

Author

Zehua Yu – *MOE Key Laboratory of Hydraulic Machinery Transients, School of power and Mechanical Engineering, Wuhan University, Wuhan 430072, China*

Yongshun Ren – *MOE Key Laboratory of Hydraulic Machinery Transients, School of power and Mechanical Engineering, Wuhan University, Wuhan 430072, China*

**Notes**

The authors declare no competing financial interest

**Acknowledgement**

This work was supported by the National Key R&D Program of China (2021YFA1202400), National Natural Science Foundation of China (51976141, 62161160311). The authors appreciate the support of the Supercomputing Center of Wuhan University for calculations.

TOC Graphic

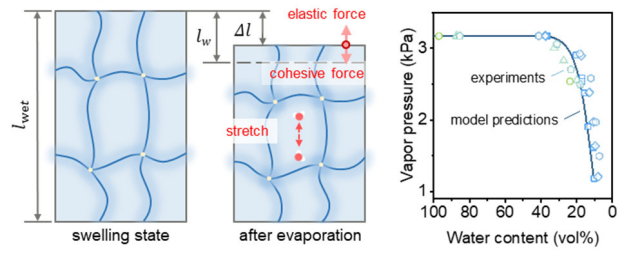

Supporting Information

# Modeling for Drying of Low-water-content Hydrogels from Aspect of Mechanical Analytics

Zehua Yu[1], Yongshun Ren[1], Kang Liu[1, †]

[1] MOE Key Laboratory of Hydraulic Machinery Transients, School of Power and Mechanical Engineering, Wuhan University, Wuhan 430072, China

**Contents**

Note S1. Fabrication of Hydrogels.

Note S2. Characterization.

Note S3. Evaporation Model.

Note S4. Hydrogel Elasticity Model.

Figure S1. Negative Pressure of Liquid Water Under Varied Stretch Ratio.

Figure S2. Mass and Thickness Ratio of Swelling/Dry Hydrogels.

Figure S3. Nanoindentation and Tensile Test of Hydrogels.

Figure S4. Elastic Modulus Fitted by Flory-Rehner statistical model.

Figure S5. Elastic Modulus Predicted by Simple Affine Network model.

Figure S6. Raman Characterization of Water in Hydrogels.

Figure S7. Elastic Modulus and Vapor Pressure of PAAM Hydrogels.

Figure S8. Deformations of Hydrogels at Various Water Contents.

**Note S1. Fabrication of Hydrogels.**

In fabrication of PAAM hydrogels, acrylamide (AAm) was used as reaction monomers; N, N′-methylenebis(acrylamide) (MBAA) was used as the crosslinking agent; 2-Hydroxy-4′-(2-hydroxyethoxy)-2-methylpropiophenone (2959) was used as the photo initiator. The concentrations of AAm were determined to be 2, 10, and 15 mol L$^{-1}$ (designated as 2M, 10M and 15M, respectively), with a fixed molar ratio of monomer to crosslinker (or photoinitiator) at 1000:1. After preparing the hydrogel precursor, it was thoroughly degassed in a vacuum chamber, cast into a mold, sealed with a transparent PET cover plate, and irradiated with 365 nm ultraviolet light until gelation occurred. The fabrication process for PHEMA hydrogels followed procedures established in previous work.[1]

**Note S2. Characterizations.**

The elastic modulus of hydrogels in completely dehydrated state was measured through nanoindentation test (NanoTest™ Vantage™). To achieve complete drying, the hydrogels were vacuum-dried at 100 °C for 48 hours. A Berkovich indenter was then used to apply a fixed load of 50 mN over a duration of 20 seconds. For hydrated hydrogels, the modulus was primarily determined through tensile tests. The hydrogels were fully swollen in water, cut into standard spindle shapes, and their water content was adjusted through controlled evaporation. During the evaporation process, the lateral surfaces of the samples were restricted, resulting in volume shrinkage primarily through thickness reduction. Tensile tests were subsequently performed at a rate of 5 mm min$^{-1}$ until fracture. The elastic modulus was calculated by linear fitting within a strain range of 0.2% to 0.8%. The same water content control methodology was used to prepare samples for Raman spectroscopy (alpha300 R). Multi-peak fitting was conducted in the range of 2800–4000 cm$^{-1}$, with the peak near 3200 cm$^{-1}$ corresponding to bound and intermediate water, and the peak at 3450 cm$^{-1}$ indicating free water.

**Note S3. Evaporation model.**

In our model, we assume that the cohesive force within the water is balanced with the elastic force generated by the hydrogel framework when the hydrogel is equilibrium in the environment. According to the cohesion-tension hypothesis, water undergoes stretching in a confined space would exhibit an absolute negative pressure. Thus, we employee negative pressure to quantitatively express the magnitude of cohesion in the stretched state of water. In previous work,[2] we calculated the negative pressure of liquid water in stretching state through molecular dynamics simulations. In the simulations, we reduced the number of water molecules in a box with fixed volume, and made the remaining water fill the box and undergoes a stretching state. We defined the stretching ratio ($\varepsilon_s$) of water as,

$$\varepsilon_s = \frac{N}{N - \Delta N} \tag{S1}$$

Here, $\Delta N$ represents the number of removed water molecules, and $N$ denotes the initial number of water molecules. The negative pressure of liquid water at various stretching extents is shown in Figure S1. By fitting these data, we constructed an equation of state for water under stretching. And the negative pressure of water ($p_{l,gel}$) can be calculated as follows,

$$p_{l,gel} = f(\varepsilon_s) \tag{S2}$$

Similarly, we defined the stretch ratio of water in hydrogels,

$$\varepsilon_s = \frac{l_{w,o} - \Delta l}{l_{w,o} - l_w} \tag{S3}$$

Note that our model assumes a one-dimensional configuration, which is appropriate for thin hydrogel layers. Therefore, the thickness ratio can be treated as equivalent to the volume ratio. Where $l_{w,o} = l_{wet} - l_{dry}$ indicates the thickness change of the hydrogel due to water evaporation, where the hydrogel transitions from a fully swollen to a fully dehydrated state. $\Delta l$ indicates the actual thickness change of the hydrogel at any water content, and $l_w = V_w / A$, where $V_w$ is the volume of evaporated water, and $A$ is the total area of the hydrogel. The value of $l_w$ can also be calculated by,

$$l_w = \frac{m_{wet} - m_t}{\rho_w A} = \frac{l_{w,o} \cdot (m_{wet} - m_t)}{l_{w,o} \rho_w A} = \frac{l_{w,o} \cdot \left[(m_{wet} - m_{dry}) + (m_t - m_{dry})\right]}{m_{wet} - m_{dry}} \tag{S4}$$

Where $m_{wet}$ is the mass of hydrogel at fully swelling state, $m_t$ is the mass of hydrogel at any water content, $m_{dry}$ is the mass of hydrogel at completely dehydrate state, $\rho_w$ is the density of water. Then we define the relative water content $\omega$ as,

$$\omega = \frac{m_t - m_{dry}}{m_{wet} - m_{dry}} \tag{S5}$$

Substituting S4, S5 into S3, we get the stretch ratio as,

$$\varepsilon_s = \frac{1 - \Delta l / l_{w,o}}{\omega} \tag{S6}$$

Then, the pressure of water in the hydrogel is obtained by substituting S6 to S2.

The force that balances the cohesive force is the elastic force generated by the hydrogel network, which can be calculated as the product of stress and strain. However, as the modulus of the hydrogel $E(\varphi_w)$ gradually increases during dehydration, the total elastic force produced by the hydrogel should be represented by the integral of the modulus-strain product over the entire strain range,

$$p_{l.gel}(\varepsilon_s) - p_e = -\int_0^{\Delta l} E(\varphi_w) \cdot dl / l_{wet} \tag{S7}$$

By combining equations S7 and S2, we can calculate both the pressure of water within the hydrogels and the total strain of the hydrogels (Figure S8).

**Note S4. Hydrogel elasticity model.**

In calculating the elastic forces of hydrogels, it is necessary to fit the modulus at varying water contents. We initially employed the classic Flory–Rehner statistical model, where the modulus $E_{gel} \propto f_2^{1/3}$.[3] The fitting results only demonstrate satisfactory performance at water contents exceeding 40 vol%, but fail to capture the sharp increase in modulus observed at water contents below 40 vol% (Figure. S4). Consequently, we shifted our focus to the Simple Affine Network model proposed by Li et al.[4] In this model, the modulus of the hydrogel is determined by the product of the elasticity per strand and the density of the strands, expressed as:

$$E_{gel} = \frac{\upsilon_{e,o} k_B T}{\lambda_s} \frac{3+\alpha^4}{(1-\alpha^2)^2} \tag{S8}$$

Here, $\upsilon_{e,o}$ denotes the strand density in the reference state, $k_B$ is the Boltzmann constant, and $\lambda_s$ represents the volume expansion ratio, with $\alpha = \lambda_s \cdot R_o / R_{max}$, where $R_o$ is the strand length in the reference state and $R_{max}$ is the length in the fully swollen state. The reference state is assumed to the state of hydrogels when just synthesized. The fitting results are similar to that of the Flory–Rehner model, and the capture of the sharp increase in modulus at low water content is still lacked (Figure S5).

To address this limitation, we developed a hydrogel spring model, assuming that the hydrogel can be represented by three springs in series, corresponding to the polymer network, bound/intermediate water, and free water. The modulus of the polymer network reflects that of the fully dehydrated hydrogel, while free water contributes negligible tensile strength and is approximated by the modulus of the hydrogel in a fully swelling state. According to the series spring model, the overall elastic coefficient ($k_{gel}$) of the hydrogel is expressed as:

$$k_{gel} = \cfrac{1}{\cfrac{1}{k_p} + \cfrac{1}{k_b} + \cfrac{1}{k_f}} \tag{S9}$$

$k_p$, $k_b$ and $k_f$ are the elastic coefficient of polymer network, bound water and free water, respectively. The relationship between the elastic coefficient and the modulus is $k=E \cdot A/l$, and the thickness ratio is replaced by the volume ratio due to the one-dimensional model, so the overall modulus can be expressed as

$$E_{gel} = \cfrac{1}{\cfrac{\varphi_p}{E_p} + \cfrac{\varphi_b}{E_b} + \cfrac{\varphi_f}{E_f}} \tag{S10}$$

Since the areas of polymer network, bound water and free water are identical, A cancels out on both sides of the equation. At lower water content, only bound water is present within the hydrogel network, exhibiting a high modulus due to strong interactions with the polymer matrix. As water content increases, intermediate water begins to appear, adding hydration layers around polymer chains and sharply increasing water mobility. In addition, the hydrogen bond belongs to the intermediate water is less strong than that of the bound water. Therefore, we describe the $E_b$ as a variable. In the formula, $E_b(\varphi_b)=E_{b,o} \cdot \exp(\alpha \cdot \varphi_b)$. Where $\varphi_w=\varphi_b+\varphi_f$. And after free water appears, the $E_b$ was set as a constant value at the $\varphi_b$. The results showed that our model fitted well with experimental data (Figure 3 and Figure S7).

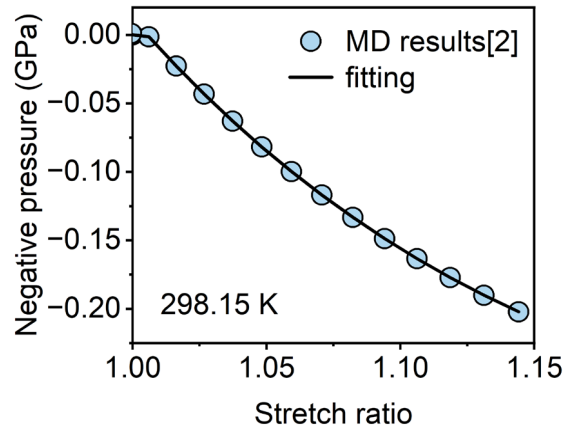

*Figure S1.* Relationship between absolute negative pressure and stretch ratio of liquid water. Stretch ratio of 1 represents liquid water under standard conditions. When stretch ratio is between 1 and 1.006, linear fitting is used to fit the data. When stretch ratio is between 1.006 and 1.15, quadratic polynomial is utilized to fit the data.

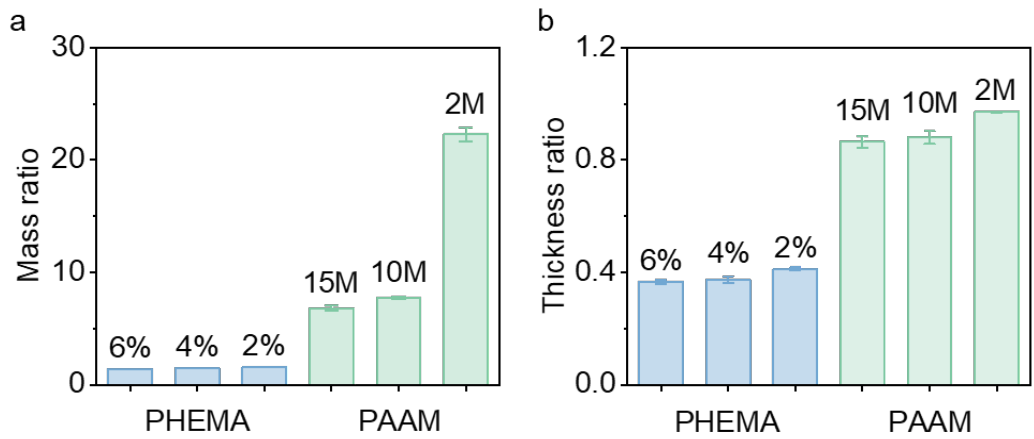

*Figure S2.* Mass ratio (a) and thickness ratio (b) of complete swelling hydrogels and dry hydrogels for various PHEMA and PAAM hydrogels. Since we constrained the shrinkage of the hydrogel in the plane direction during the thickness test, the thickness change also represents the volume change.

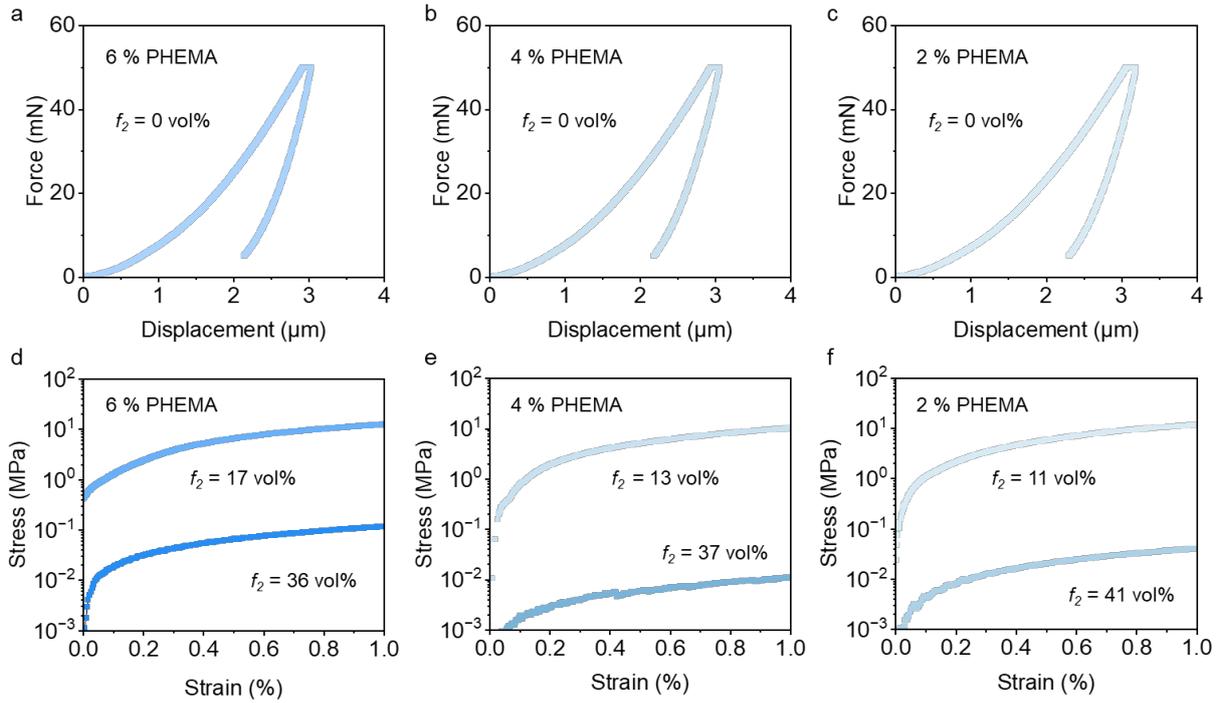

*Figure S3.* Mechanical properties characterization of hydrogels. (a-c) Nanoindentation tests for (a) 6%, (b) 4%, (c) 2% PHEMA hydrogels under water contents of 0 vol%. (d-f) Tensile tests for (d) 6%, (e) 4%, (f) 2% PHEMA hydrogels under fully swelling state and modulus transition stage.

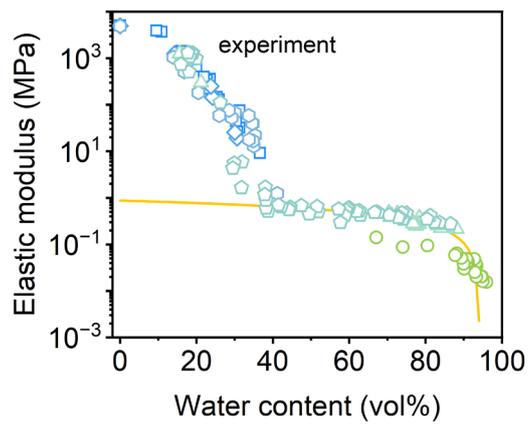

*Figure S4.* Elastic modulus of hydrogels from experiment and the fitting results based on the Flory-Rehner statistical model.

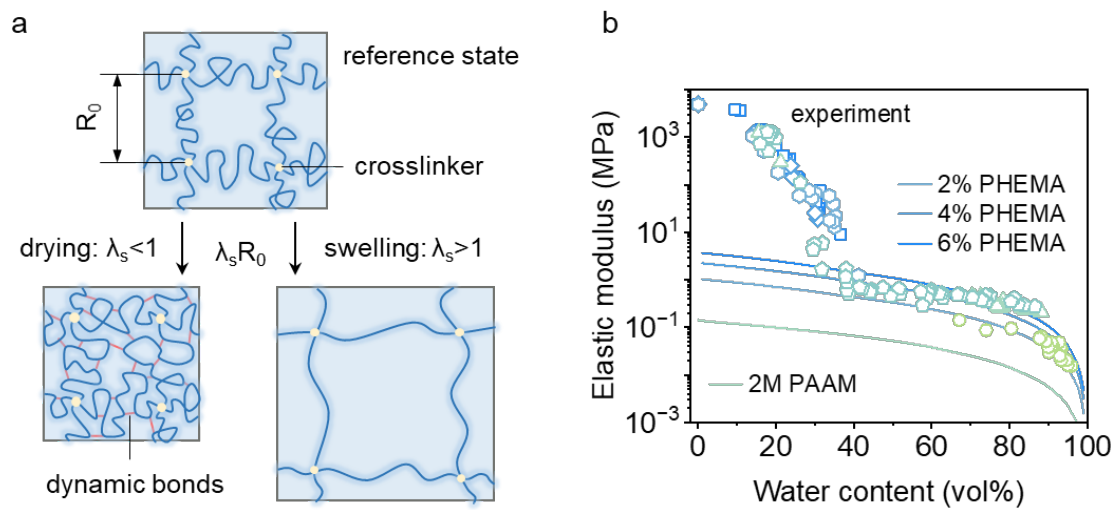

*Figure S5.* Elastic modulus of hydrogel from experiment and the modulus predicted by the Simple Affine Network model. (a) Schematic of the Simple Affine Network model. (b) Modulus predicted by the Simple Affine Network model.

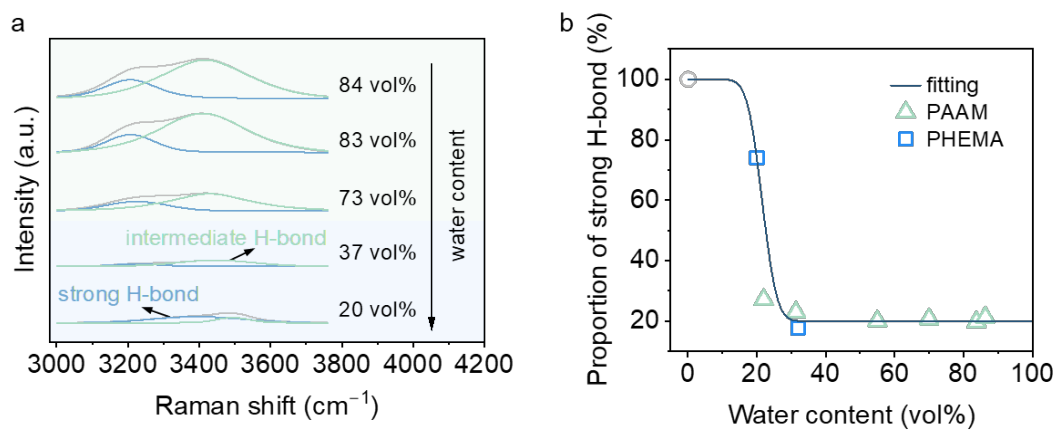

*Figure S6.* Raman characterization of water in hydrogels. (a) Raman characterization of water. The green shade for PAAM and the blue shade for PHEMA. The peak at about 3200 $cm^{-1}$ represents the strong hydrogen bonds while the peak at about 3450 $cm^{-1}$ represents the intermediate hydrogen bonds. (b) Proportion of the water with strong hydrogen bonds in total water within hydrogels. The water with strong hydrogen bonds is recognized as bound/intermediate water.

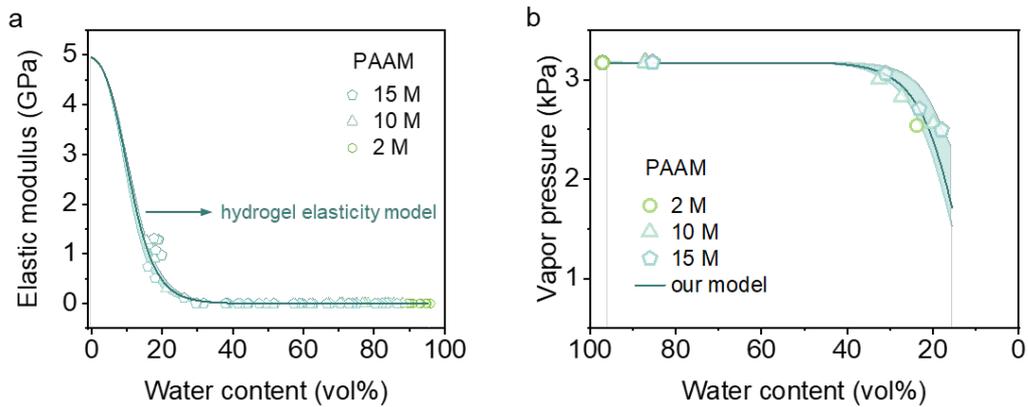

*Figure S7.* Elastic Modulus (a) and Vapor Pressure (b) of PAAM Hydrogels. The green line indicates the results from the hydrogel elasticity model (a) and results from the predictions of the evaporation model (b).

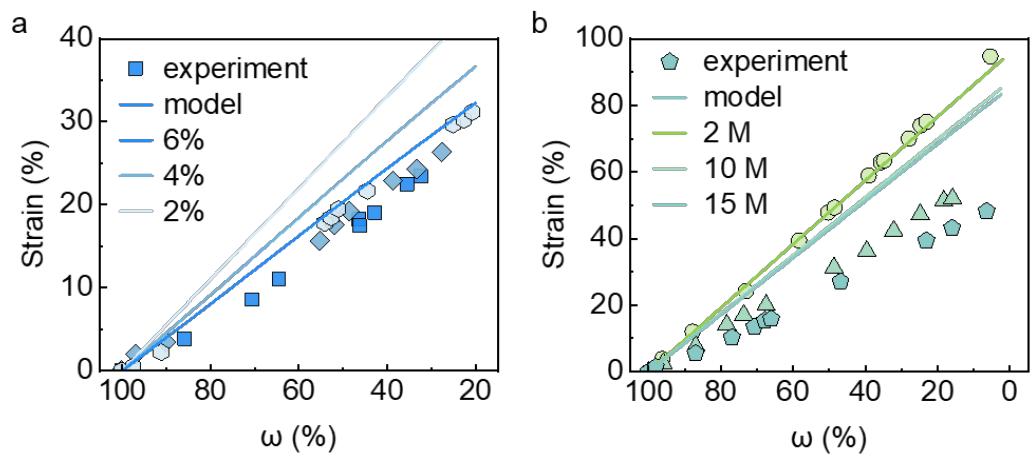

*Figure S8.* Total deformation of hydrogels at various water contents. The hydrogels are equilibrium at different environment humidity conditions. (a) PHEMA and (b) PAAM hydrogels.